%% file: draft.tex
\documentclass[prd,amsmath,amssymb,showpacs,superscriptaddress,nofootinbib,twocolumn]{revtex4-1}
\usepackage{graphicx}
\usepackage{epsfig,graphics,subfigure,psfrag,amsmath,amssymb}
\usepackage{lineno}
\usepackage{dcolumn}
\usepackage{bm}
\usepackage{overpic}
\usepackage{xspace}
\usepackage{xcolor}
\usepackage{rotating}
\usepackage{color}
\usepackage{multirow}
\usepackage{colortbl}
\usepackage{epstopdf}
\usepackage{float}
\usepackage{makecell}
\usepackage[colorlinks,linkcolor=blue,anchorcolor=blue,citecolor=blue]{hyperref}
\newcommand{\jpsi}{J/\psi}

\newcommand{\etap}{\eta^{\prime}}

\newcommand{\BESIIIorcid}[1]{\href{https://orcid.org/#1}{\hspace*{0.1em}\raisebox{-0.45ex}{\includegraphics[width=1em]{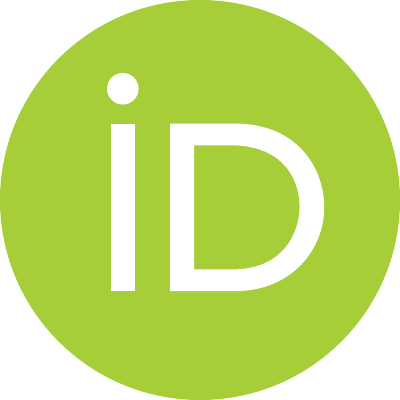}}}}

\newcommand{\BESIII}{BES\uppercase\expandafter{\romannumeral3}\xspace}

\begin{document}
\title{\boldmath An improved measurement of $\eta^\prime\rightarrow e^{+}e^{-}\omega$}
\author{
\begin{small}
  \begin{center}
    \input{authorlist_2025-12-30}
\end{center}
\end{small}
}

\date{\today}

\begin{abstract}
Using a sample of $(10087 \pm 44) \times 10^{6}$ $J/\psi$ events collected with the BESIII detector, an improved measurement of the decay $\eta^{\prime}\rightarrow e^{+}e^{-}\omega$, with $\omega\rightarrow\pi^{+}\pi^{-}\pi^{0}$ and $\pi^{0}\rightarrow\gamma\gamma$ is performed. The branching fraction is determined to be $\mathcal{B}(\eta^{\prime}\rightarrow e^{+}e^{-}\omega) = (1.79 \pm 0.09 \pm 0.12) \times 10^{-4}$, where the first uncertainty is statistical and the second is systematic. This result is consistent with the previous measurement and is obtained with significantly improved precision. Furthermore, the first measurement of the transition form factor cutoff parameter for this decay is reported, with $\Lambda^{-1} = (2.92 \pm 0.83 \pm 0.15)~\text{GeV}^{-1}$. These measurements provide valuable input for understanding the internal structure of the $\eta^{\prime}$ meson and testing theoretical models.

\end{abstract}

\maketitle
\section{INTRODUCTION}

The pseudoscalar meson $\eta^{\prime}$, discovered in 1964~\cite{Kalbfleisch:1964zz,Goldberg:1964zza}, remains a subject of significant interest because its decay properties serve as a probe of meson internal structure and a testing ground for low-energy quantum chromodynamics (QCD). Its decay properties can be classified
into two main categories.
Hadronic decays, such as $\eta^{\prime}\rightarrow\eta\pi\pi$, are dominant and governed by isospin conservation.
Radiative decays into vector mesons, such as $\eta^{\prime}\rightarrow\rho^{0}\gamma$ and $\eta^{\prime}\rightarrow\omega\gamma$, serve as a testing ground for model-dependent approaches, including Vector Meson Dominance (VMD) and chiral perturbation theory.

Decays of the form $\eta^{\prime}\rightarrow V e^{+}e^{-}$ (where V denotes a vector meson) are particularly insightful, as they proceed via a virtual photon $\eta^{\prime}\rightarrow V\gamma^{*}\rightarrow V e^{+}e^{-}$. The invariant mass spectrum of the $e^{+}e^{-}$ pair provides direct access to the $\eta^{\prime}$ transition form factor (TFF). The TFF encapsulates the meson's internal structure and its dependence on the momentum transfer, $q^{2}$, making it a valuable tool for phenomenological studies.
A decay of similar nature, $\eta^{\prime}\rightarrow\pi^{+}\pi^{-}e^{+}e^{-}$, has been previously measured~\cite{BESIII:2013tjj} and agrees with the theoretical predictions~\cite{Faessler:1999de}.

The branching fraction(BF) for $\eta^{\prime}\rightarrow e^{+}e^{-}\omega$ was previously measured by BESIII to be $\mathcal{B}(\eta^{\prime}\rightarrow e^{+}e^{-}\omega) = (1.97 \pm 0.34 \text{ (stat.)} \pm 0.17 \text{ (syst.)}) \times 10^{-4}$ using $(1310.6 \pm 7.2)\times 10^6$ $J/\psi$ events~\cite{BESIII:2015jiz}.

In this work, we present an improved analysis of the decay $\eta^{\prime}\rightarrow e^{+}e^{-}\omega$, with $\omega\rightarrow\pi^{+}\pi^{-}\pi^{0}$ and $\pi^{0}\rightarrow\gamma\gamma$, using the process $J/\psi\rightarrow\gamma\eta^{\prime}$. The analysis utilizes a significantly larger data sample of $(10087 \pm 44) \times 10^{6}$ $J/\psi$ events~\cite{BESIII_2021_njpsi}. We report a more precise measurement of the BF and, for the first time, a measurement of the
TFF parameter for this decay channel.

\section{BESIII DETECTOR AND MONTE CARLO SIMULATION}
The BESIII detector~\cite{BESIII_2009_detector} records symmetric $e^+e^-$ collisions
provided by the BEPCII storage ring~\cite{Yu:IPAC2016-TUYA01}
in the center-of-mass energy range from 1.84 to 4.95~GeV,
with a peak luminosity of $1.1 \times 10^{33}\;\text{cm}^{-2}\text{s}^{-1}$
achieved at $\sqrt{s} = 3.773\;\text{GeV}$.
The BESIII detector has accumulated the world's largest sample of $J/\psi$ events in this energy region , which provides the foundation for this high-precision study~\cite{BESIII:2020nme}. The cylindrical core of the BESIII detector covers 93\% of the full solid angle and consists of a helium-based multilayer drift chamber~(MDC), a plastic scintillator time-of-flight
system~(TOF), and a CsI~(Tl) electromagnetic calorimeter~(EMC),
which are all enclosed in a superconducting solenoidal magnet
providing a 1.0~T magnetic field. The solenoid's magnetic field was 0.9~T during the data-taking period in 2012, which accounts for approximately 10\% of the total $J/\psi$ data sample.
The solenoid is supported by an
octagonal flux-return yoke with resistive plate counter muon
identification modules interleaved with steel.
The charged-particle momentum resolution at $1~{\rm GeV}/c$ is
$0.5\%$, and the
${\rm d}E/{\rm d}x$
resolution is $6\%$ for electrons
from Bhabha scattering. The EMC measures photon energies with a
resolution of $2.5\%$ ($5\%$) at $1$~GeV in the barrel (end cap)
region. The time resolution in the TOF barrel region is 68~ps, while
that in the end cap region was 110~ps.
The end cap TOF
system was upgraded in 2015 using multigap resistive plate chamber
technology, providing a time resolution of
60~ps~\cite{BESIII_2019_detector}, which benefits 87$\%$ of the data used in this analysis.

Simulated data samples produced with a  {\sc geant}4-based~\cite{GEANT4:2002zbu} Monte Carlo (MC) package, 
including the geometric description of the BESIII detector and the detector response, are used to determine detection efficiencies and to estimate backgrounds. The simulation models the beam energy spread and initial state radiation (ISR) in the $e^+e^-$ annihilations with the generator {\sc kkmc}~\cite{Jadach:2000ir,Jadach:1999vf}. All particle decays are modeled with {\sc evtgen}~\cite{Lange:2001uf,Ping:2008zz} using BFs either taken from the Particle Data Group (PDG)~\cite{ParticleDataGroup:2024cfk}, when available, or otherwise estimated with {\sc lundcharm}~\cite{Chen:2000tv,Yang:2014vra}.  Final state radiation (FSR) from charged final state particles is incorporated using the {\sc photos} package~\cite{Richter-Was:1992hxq}.

The decay chain of the signal process $J/\psi\to\gamma\eta^{\prime}$, $\eta^{\prime}\to e^+e^-\omega$, $\omega\to\pi^+\pi^-\pi^0$, and $\pi^0\to\gamma\gamma$ containing 5 million events are produced. 
In this analysis, the simulation of the $\eta^{\prime}\to e^+e^-\omega$ process uses the parameters of the TFF,
which is parametrized following the approach in Ref.~\cite{Qin:2017vkw}.
Within the effective meson theory framework, the squared amplitude for the decay $\eta^\prime\rightarrow e^{+}e^{-}\omega$ is given by
\begin{align}
\label{eq_a}
    \vert \mathcal{A}_{\etap\rightarrow e^{+}e^{-}\omega} \vert^2
    &= \frac{2^6\pi^2M^3_{\etap}\alpha\Gamma_{\etap\rightarrow \gamma \omega}}{(M^2_{\etap}-M^2_{\omega})^3}\vert F(q^2)\vert^2  \notag \\
    & \cdot \frac{(M^2_{\etap}-q^2-M^2_{\omega})^2 - 4M^2_{\omega}q^2}{q^2}(2-\beta^2\sin^2{\theta}),
\end{align}
where $\alpha=1/137$, $q^2 = (q_{e^+} + q_{e^-})^2$, $q_{e^\pm}$ denotes the four-momentum of $e^{\pm}$; $\mathrm{\Gamma}_{\etap\rightarrow \gamma \omega}$ denotes the decay width of $\eta^\prime\rightarrow \gamma \omega$ and is quoted from the PDG~\cite{ParticleDataGroup:2024cfk}; $M_{\etap}$ and $M_{\omega}$ are the masses of the $\etap$ and $\omega$, respectively, quoted from the PDG~\cite{ParticleDataGroup:2024cfk};
$\beta = \sqrt{1-\frac{4M^2_{e^{\pm}}}{q^2}}$; and $\theta$ is the polar angle of $q_{e^\pm}$ in the $e^{+}e^{-}$ rest frame with respect to the direction of the flight of the $q_{e^+}q_{e^-}$ in the pseudoscalar rest frame. In the VMD model, the interactions between a virtual photon and hadrons are assumed to be dominated by a superposition of neutral vector meson states. The TFF can be
parametrized as~\cite{Landsberg:1985gaz}
\begin{align}
F(q^{2})=N\sum\frac{g_{\eta^{\prime}\gamma \omega}}{2g_{\omega\gamma}}\frac{m_{\omega}^2}{m_{\omega}^2-q^2-i\Gamma_{\omega}m_{\omega}},
\end{align}
where $N$ is a normalization constant ensuring $F(0)=1$;  $m_{\omega}$ and $\Gamma_{\omega}$ are the mass and width of $\omega$, and $g_{\eta^{\prime}\gamma \omega}$ and $g_{\omega\gamma}$ correspond to the respective coupling constants. When there is a single dominant vector meson, the single-pole approximation is often used:
\begin{align}
F(q^{2})=\frac{1}{1-{q^{2}}/{\Lambda^{2}}}.
\end{align}

Here, the single parameter $\Lambda$ can be experimentally determined from the slope of the TFF, defined as:
\begin{equation}
\mathrm{slope}\equiv \frac{d F}{dq^{2}}\bigg|_{q^2=0}=\frac{1}{\Lambda^2}.
\end{equation}

\section{Event selection and background analysis}

Charged tracks
with a polar angle $|\cos \theta| \leq 0.93$ and passing the interaction point within $\pm 20$ cm
along the beam direction and within $\pm 2$ cm in the plane perpendicular to the beam are accepted.
Photon candidates are required to have an energy deposition above $25~\mathrm{MeV}$ in the barrel EMC ($|\cos \theta| < 0.8$) or $50~\mathrm{MeV}$ in the end cap EMC ($0.86 < |\cos \theta|< 0.92$). To exclude showers from charged particles, the opening angle between the shower direction and the charged tracks extrapolated to the EMC must be greater than 10 degrees. A requirement on the EMC timing ($0 \leq t \leq 700$ ns) is used to suppress electronic noise and energy deposits unrelated to the event.
The final state of interest is studied through the decay chain $J/\psi\rightarrow\gamma\etap, \etap\rightarrow e^+e^-\omega, \omega\to\pi^+\pi^-\pi^0, \pi^0\to\gamma\gamma$, and candidate events are required to have four good charged tracks with a net charge of zero and at least three photons. In the two-body decay of $J/\psi\to\gamma\eta'$, the radiative photon carries a unique energy of $1.4$ $\mathrm{GeV}$. Hence, to suppress background events, the most energetic photon is required to have an energy greater than 1 $\mathrm{GeV}$.
The remaining two photons are used to form $\pi^0$ candidates. The $\omega$ candidates are reconstructed by combining two oppositely charged pion tracks with the $\pi^0$, and $\eta^{\prime}$ candidates are formed by combining
the pair of $e^+e^-$ with the $\omega$.

The combined information of the specific ionization energy loss (${\rm d}E/{\rm d}x$) measurements in the MDC and the flight time measured in the TOF is used to form particle identification (PID) confidence levels for the $e$ and $\pi$ hypotheses.
To select candidate events, the combination with the smallest $\chi^2_{\rm{4C+PID}}$ is retained~\cite{BESIII:2015jiz}.
Here, $\chi^2_{\rm{4C+ PID}}=\chi^2_{\rm{4C}}+\sum_{i=1}^4\chi^2_{\rm PID}(i)$ is the sum of the chi-squares from the four-constraint (4C) kinematic fit and PID,
where $i$ denotes the index of the four charged tracks used to form an $\eta^{\prime}$ candidate.
The combination with $\chi^2_{\rm{4C+ PID}} < 70$ is kept as an $e^+ e^- \pi^+ \pi^- \gamma \gamma \gamma$ candidate.
To reject possible backgrounds with mis-identification between $e$ and $\pi$, a 4C kinematic fit is performed under the hypotheses $J/\psi \to \pi^+ \pi^- \pi^+ \pi^- \gamma \gamma \gamma$. The events with a $\chi^2_{\rm 4C}$ value for the signal hypothesis larger than that for the background hypothesis are discarded.
The selected photon with maximum energy is taken as the radiative photon.
The other two photons are further required to be consistent with a $\pi^0$ candidate,
$|M(\gamma \gamma) - M(\pi^0)| < 0.015~ \mathrm{GeV/c^2}$, where $M(\pi^0)$ is the $\pi^0$ nominal mass~\cite{ParticleDataGroup:2024cfk}.

\begin{figure}[htbp]
\centering
\begin{subfigure}{}
\includegraphics[scale=0.35]{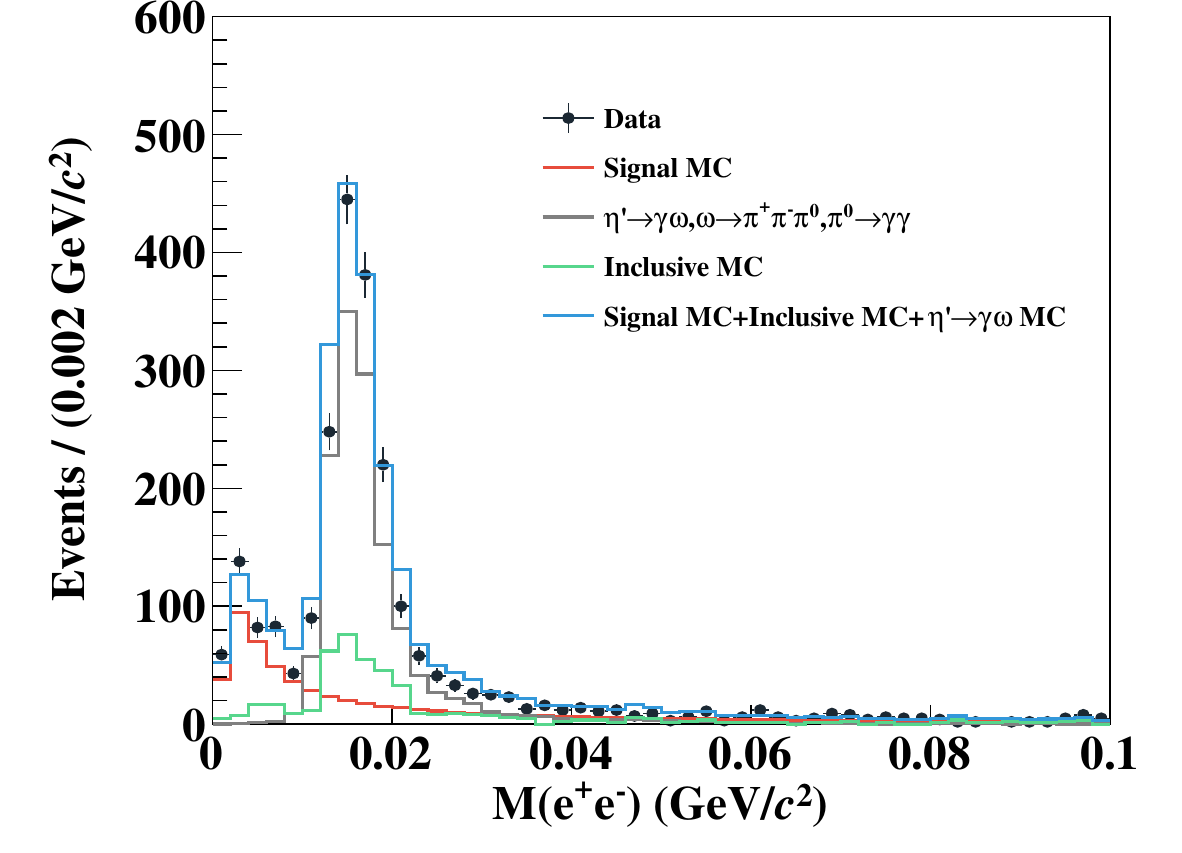}
\put(-155,115){(a)}
\end{subfigure}
\begin{subfigure}{}
\includegraphics[scale=0.35]{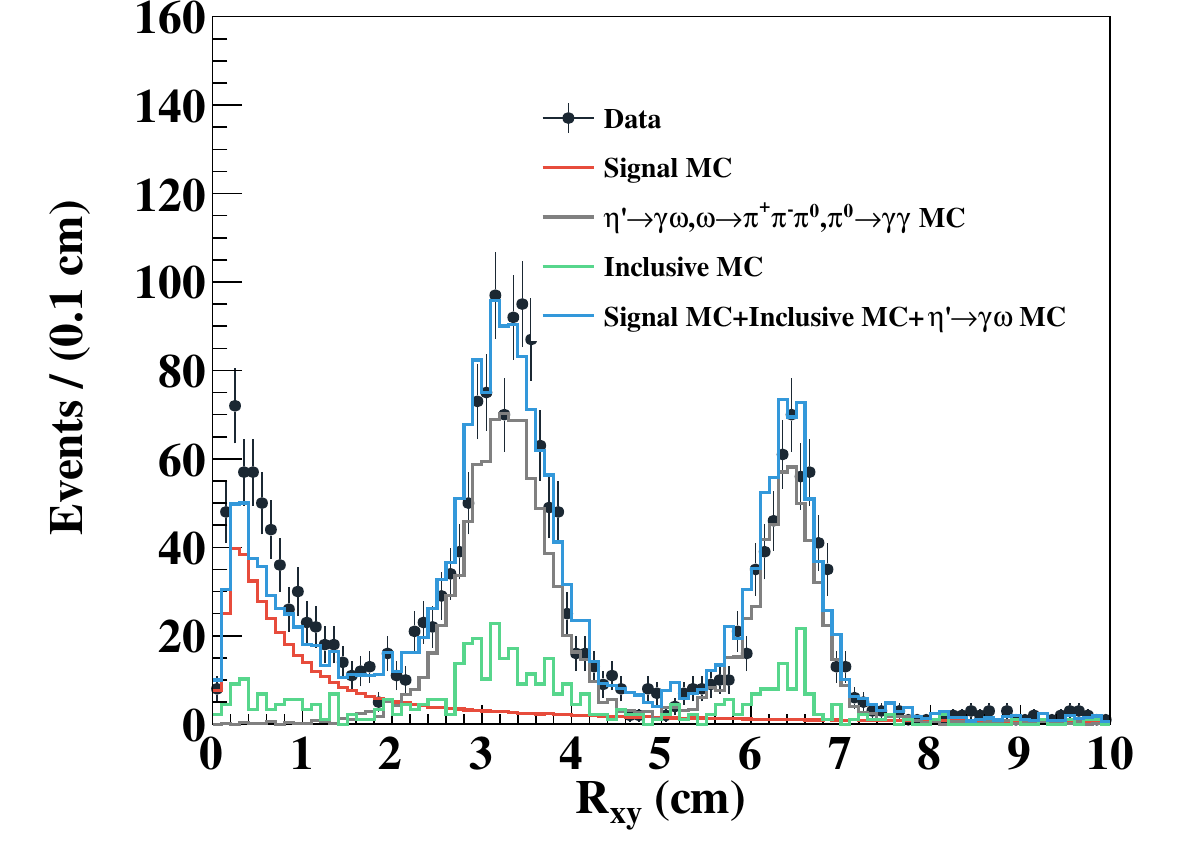}
\put(-155,115){(b)}
\end{subfigure}
\caption{
The distributions of (a) $M(e^{+}e^{-})$ and (b) the transverse distance $R_{xy}$ of the $e^{+}e^{-}$ vertex from the interaction point.
The dots with error bars represent the data, the red line is the simulated signal shape, the gray line is the $J/\psi\rightarrow\gamma\etap, \etap\rightarrow\gamma\omega, \omega\to\pi^+\pi^-\pi^{0}, \pi^0\to\gamma\gamma$ MC shape, the green line denotes the background from the inclusive MC sample,
and the blue line is the sum of all components.
}
\label{Mee_Rxy}
\end{figure}

The primary peaking background comes from $J/\psi\rightarrow\gamma\etap, \etap\rightarrow\gamma\omega, \omega\to\pi^+\pi^-\pi^0, \pi^0\to\gamma\gamma$ events, where a photon converts to an $e^+e^-$ pair at the beam pipe or the inner wall of the MDC.
For such conversion backgrounds, the invariant mass of the $e^+e^-$ pair, $M(e^+e^-)$, is expected to be close to zero. However, since the default BESIII tracking algorithm assumes the interaction point as the common origin of all tracks, the directions of tracks originating elsewhere are mis-reconstructed. As a result, the conversion pair acquires an artificial opening angle, leading to a reconstructed invariant mass larger than the true value.
Therefore, the conversion background appears as a pronounced peak at about $0.015~\mathrm{GeV}/c^2$ in the $M(e^+e^-)$ distribution, as shown in Fig.~\ref{Mee_Rxy}(a). The distribution of the distance from the reconstructed vertex of an $e^+e^-$ pair to the $z$ axis, $R_{xy}$~\cite{GamConv}, is shown in Fig.~\ref{Mee_Rxy}(b). The peaks around $R_{xy} = 3.5$~cm and $R_{xy} = 6.5$~cm correspond to the position of the beam pipe and the inner wall of the MDC, respectively.
These structures are characteristic of photon conversion backgrounds.

\begin{figure}[htbp]
\centering
\begin{subfigure}{}
\includegraphics[scale=0.35]{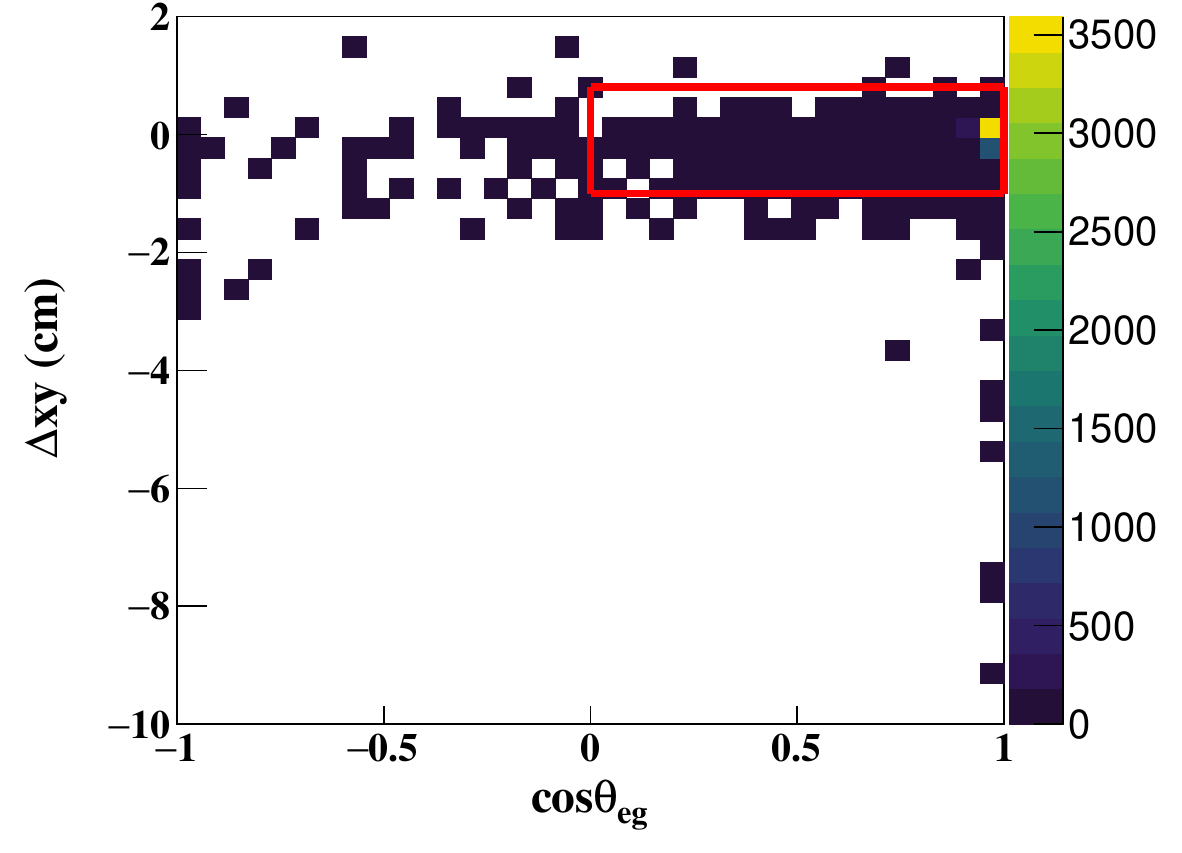}
\put(-100,60){(a) }
\end{subfigure}
\begin{subfigure}{}
\includegraphics[scale=0.35]{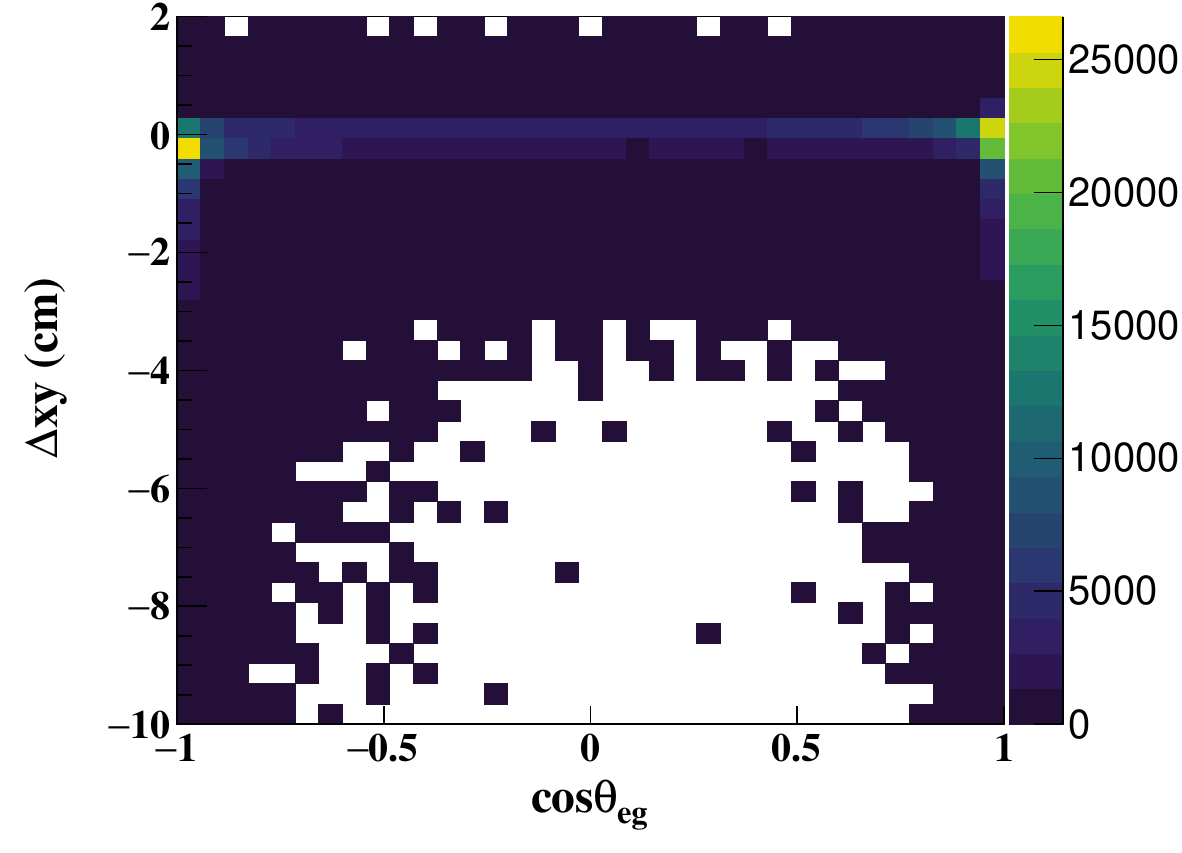}
\put(-100,50){(b)}
\end{subfigure}
\caption{The distributions of $\Delta_{xy}$ versus $\cos\theta_{eg}$ for the MC events of (a)~the background $J/\psi\to\gamma\eta^\prime, \eta^\prime\to\gamma\omega,\omega\to\pi^+\pi^-\pi^0,\pi^0\to\gamma\gamma$ and (b) the signal process. Events located in the red rectangle are rejected.}
\label{gamconv}
\end{figure}

Events with $R_{xy} < 2$ cm are retained. For events with $R_{xy} \geq 2 ~\mathrm{cm}$, those satisfying both $\cos \theta_{eg} >$ 0 and $-1.0<\Delta_{xy}< 0.8 ~\mathrm{cm}$ are excluded as shown in Fig.~\ref{gamconv}(a), while the others are retained to improve statistics.
Here, $\cos\theta_{eg}$ is defined as the cosine of the angle between the momentum of the $e^+e^-$ pair and the direction from the interaction point to the conversion point of the virtual photon $\gamma^*\to e^+e^-$.
The electron and positron tracks project to two circles in the $xy$ plane, and $\Delta_{xy}$ is the distance between the intersections of these two circles along the line connecting their centers~\cite{GamConv}.
The distributions of $\Delta_{xy}$ versus $\cos\theta_{eg}$ for the simulated signal and background MC samples are displayed in Fig.~\ref{gamconv}.

\begin{figure}[htbp]
\centering
\begin{subfigure}{}
\includegraphics[scale=0.35]{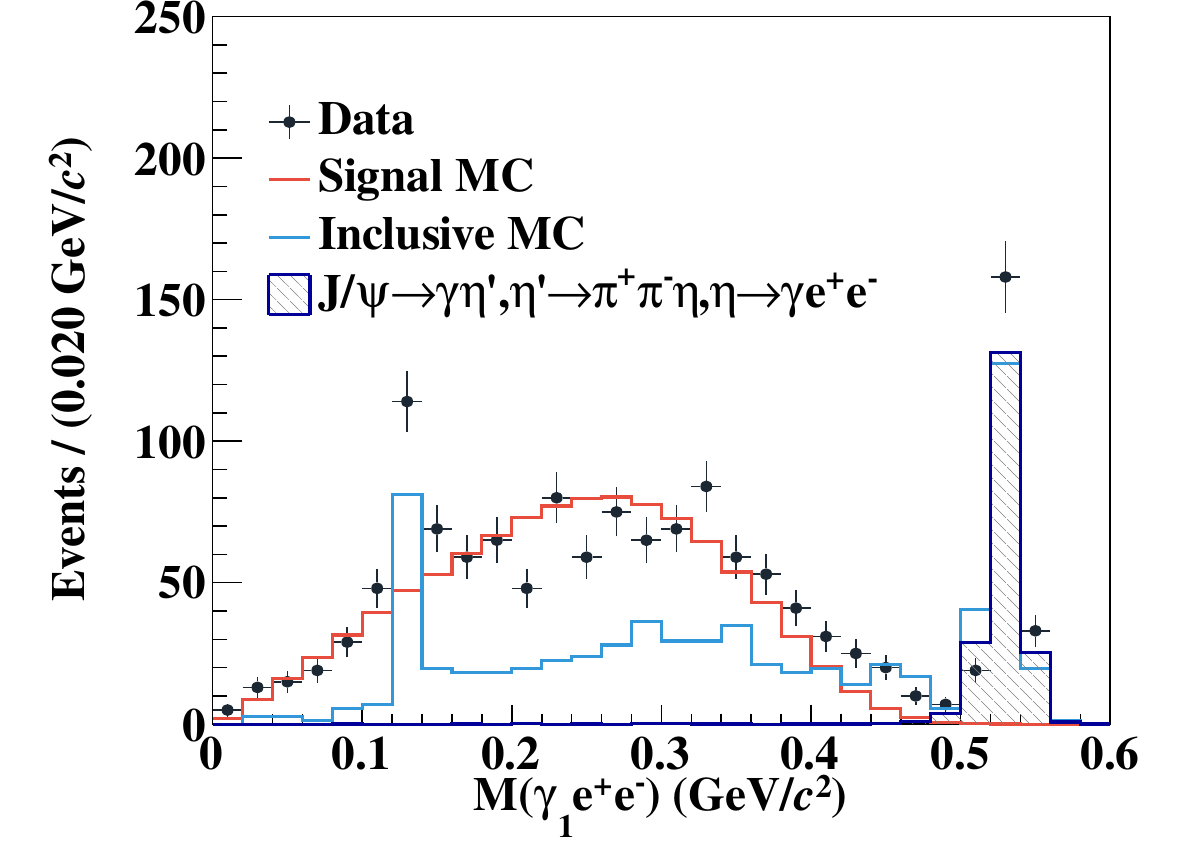}
\put(-30,110){(a)}
\end{subfigure}
\begin{subfigure}{}
\includegraphics[scale=0.35]{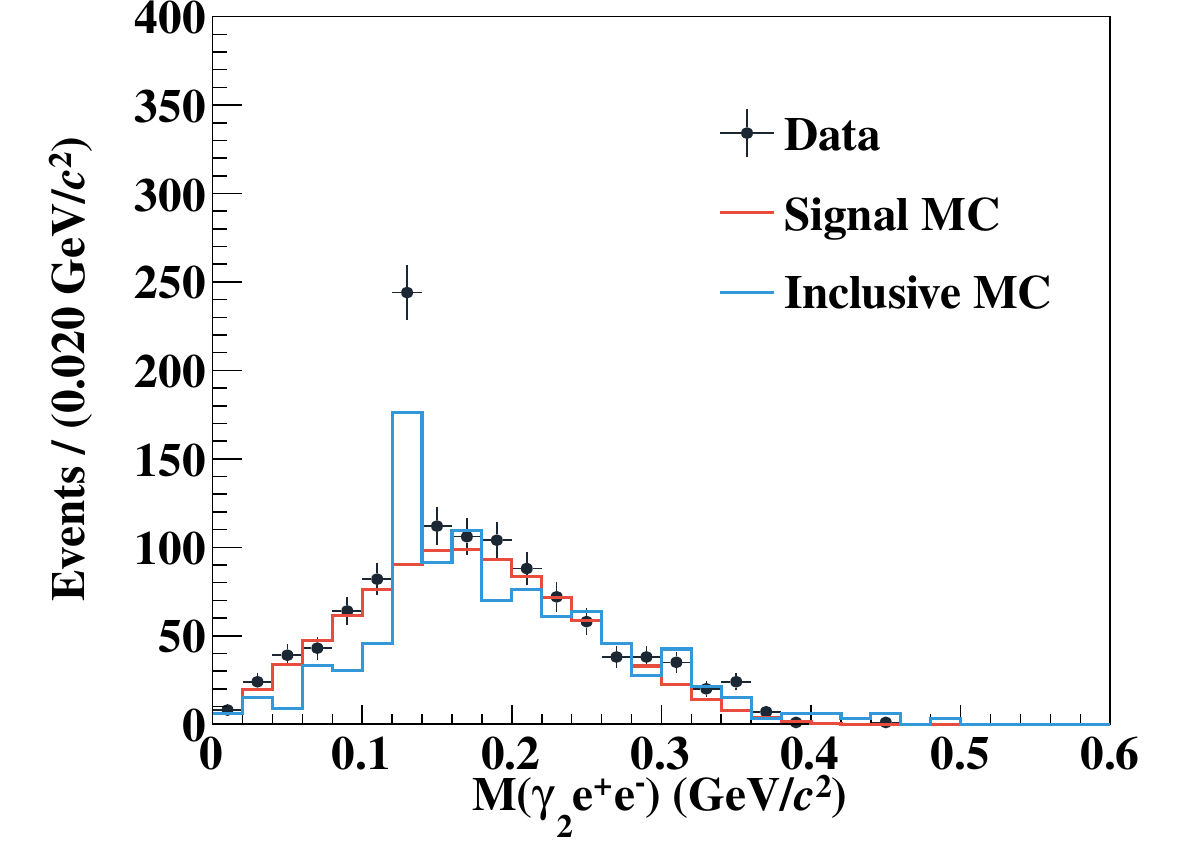}
\put(-30,110){(b)}
\end{subfigure}
\begin{subfigure}{}
\includegraphics[scale=0.35]{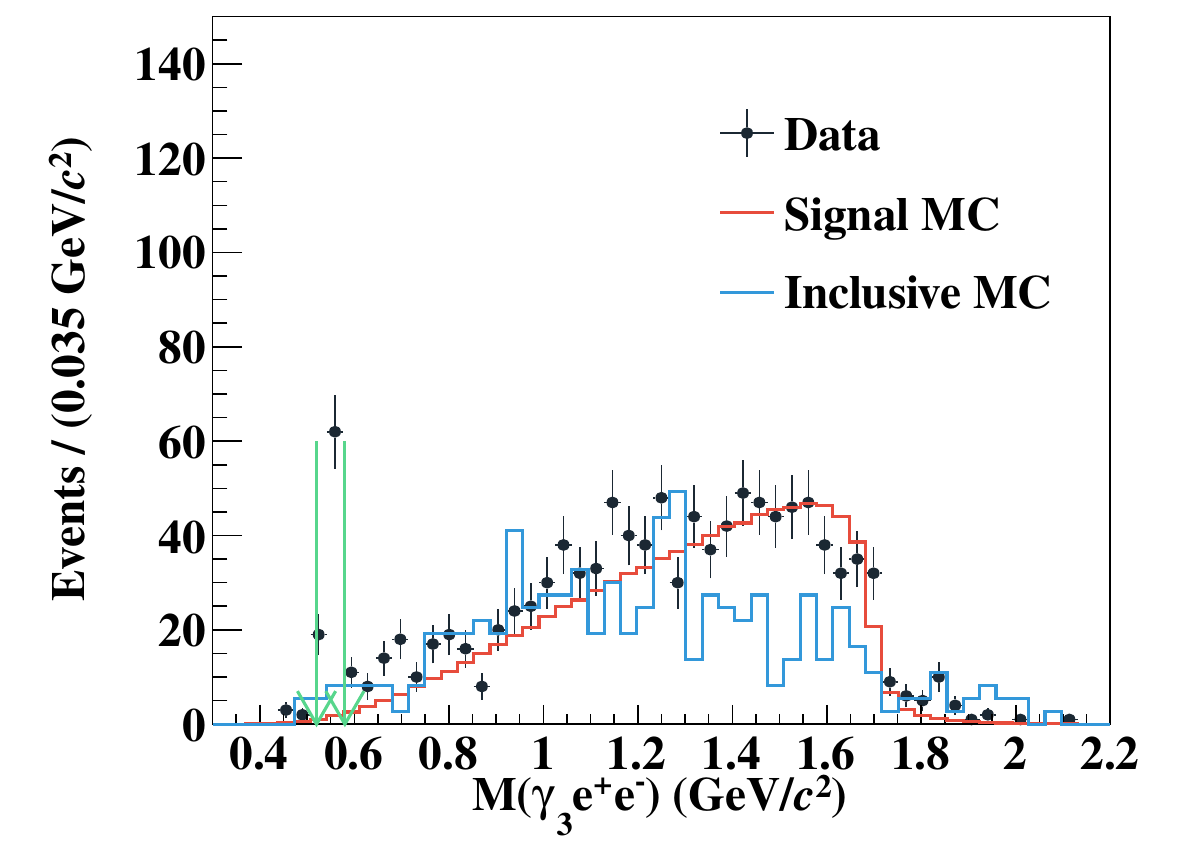}
\put(-30,110){(c)}
\end{subfigure}
\caption{The distributions of (a)~$M(\gamma_{1}e^+e^-)$, (b)~$M(\gamma_{2}e^+e^-)$ and (c)~$M(\gamma_{3}e^+e^-)$. The dots with error bars represent data, the red line represents the signal MC sample, the blue line represents the inclusive MC sample, and the gray histogram represents the background from $J/\psi\to\gamma\etap, ~\etap\to\pi^+\pi^-\eta,~\eta\to\gamma e^+e^-$. The background region for $\eta\to\gamma e^+e^-$ is indicated by the green arrows in panel (c). }
\label{gamee}
\end{figure}

To suppress potential background events from the $\pi^0\to\gamma e^+ e^-$ decay, we require events to satisfy $\left| M(\gamma_1\gamma_2)- M(\pi^0) \right|<\left|M(\gamma_{1} e^+ e^-)-M(\pi^0) \right|$, $\left| M(\gamma_1\gamma_2)- M(\pi^0) \right|<\left|M(\gamma_{2} e^+ e^-)-M(\pi^0) \right|$.
Additionally, events with $0.52 < M(\gamma_{3}e^+e^-)< 0.58$ GeV$/c^2$ are vetoed to suppress the background from the $\eta\to\gamma e^+ e^-$ process.
Here, $\gamma_1$, and $\gamma_2$ correspond to the two photons from the $\pi^0$ decay, while $\gamma_3$ denotes the radiative photon from the $\eta\to\gamma e^+ e^-$ decay.
The distributions of the invariant masses of $\gamma_1 e^+ e^-$, $\gamma_2 e^+ e^-$ and $\gamma_3 e^+ e^-$ are shown in Fig.~\ref{gamee}. To further suppress the potential background from $\jpsi\to\gamma\eta^{\prime}$, $ \eta^{\prime} \to \pi^+ \pi^- \eta$, $\eta\to\gamma e^+ e^-$, events with $M(\gamma_{1}e^+e^-) > 0.5~ \mathrm{GeV}/c^2$ are rejected, as shown in Fig.~\ref{gamee}(a).

\begin{figure}[htbp]
\centering
\begin{subfigure}{}
\includegraphics[scale=0.35]{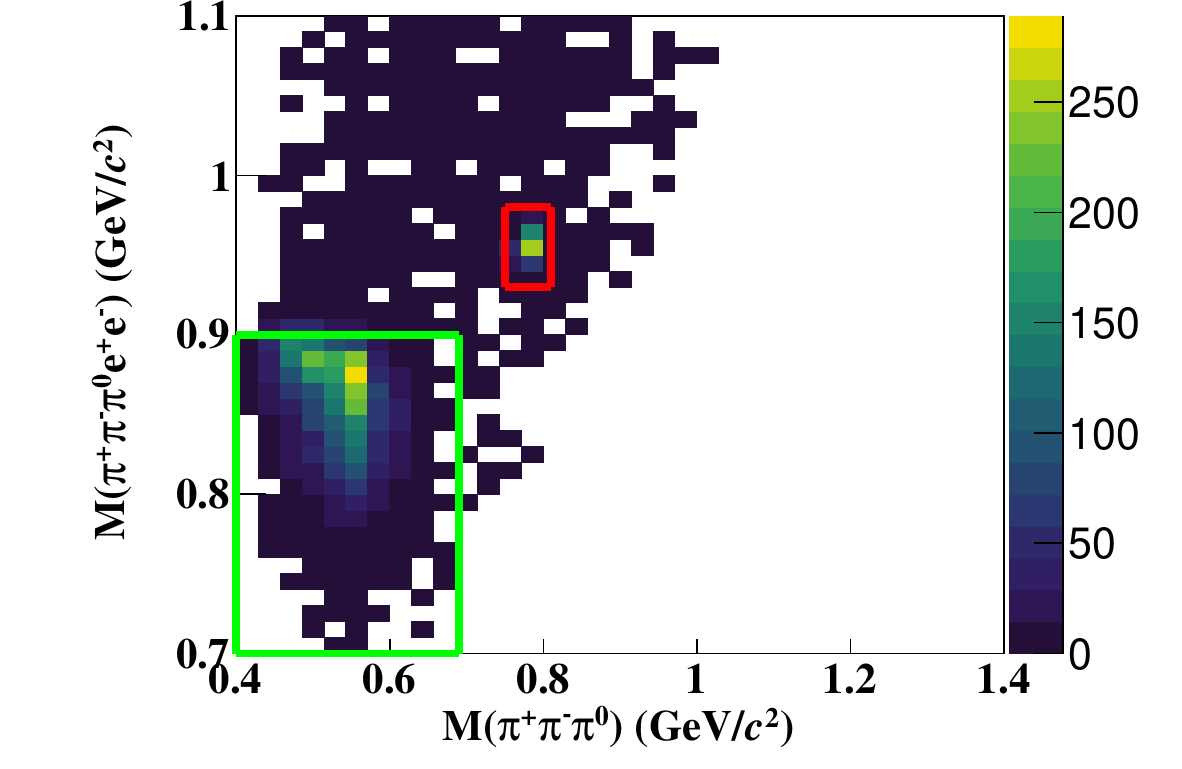}
\put(-60,100){(a)}
\end{subfigure}
\begin{subfigure}{}
\includegraphics[scale=0.35]{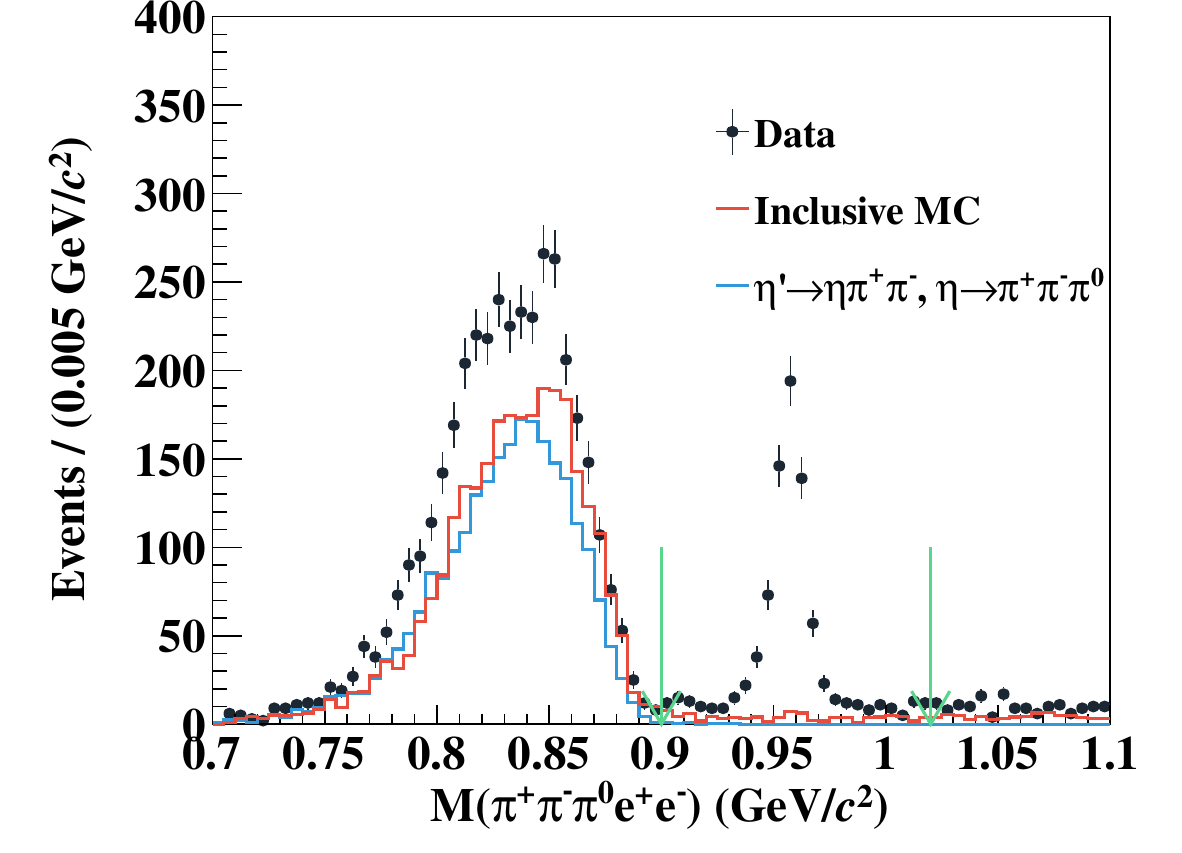}
\put(-150,110){(b)}
\end{subfigure}
\caption{(a) The distribution of $M(\pi^{+}\pi^{-}\pi^{0}e^{+}e^{-})$
  versus $M(\pi^{+}\pi^{-}\pi^{0})$ for data. The region for the
  background from $\eta^\prime\rightarrow \eta\pi^{+}\pi^{-}$,
  $\eta\rightarrow\pi^{+}\pi^{-}\pi^{0}$ is indicated by the green
  square and the signal region is indicated by the red square. (b) The
  distribution of $M(\pi^{+}\pi^{-}\pi^{0}e^{+}e^{-})$. The dots with
  error bars represent data, the red line represents the inclusive MC
  sample, and the blue line represents the background from
  $\eta^\prime\rightarrow \eta\pi^{+}\pi^{-}$,
  $\eta\rightarrow\pi^{+}\pi^{-}\pi^{0}$. The $\eta^{\prime}$ signal
  region is shown in the green arrows.}
\label{3piee}
\end{figure}

The distribution of $M(\pi^{+}\pi^{-}\pi^{0}e^{+}e^{-})$ versus $M(\pi^{+}\pi^{-}\pi^{0})$ is shown in Fig.~\ref{3piee}(a),
where a significant background contribution is observed.
Based on the study of the inclusive MC sample, the dominant contribution is from the process $\eta^\prime\rightarrow \eta\pi^{+}\pi^{-}$, $\eta\rightarrow\pi^{+}\pi^{-}\pi^{0}$, as indicated in Fig.~\ref{3piee}(b). To suppress this background, the invariant mass $M(\pi^+\pi^-\pi^0e^+e^-)$ is required to be between 0.90 $\mathrm{GeV}/c^2$ and 1.02 $\mathrm{GeV}/c^2$.

To investigate possible background contributions, the same selection criteria are applied to the inclusive MC sample of 10 billion $J/\psi$ events.
No significant peaking background is observed in the distribution of $M(\pi^+\pi^-\pi^0e^+e^-)-M(\pi^+\pi^-\pi^0)$.

An off-resonance data sample of 167.06 $\mathrm{pb}^{-1}$ taken at $\sqrt s=3.08$ $\mathrm{GeV}$ is employed to estimate the possible background contribution from the continuum process.
After applying the same selection criteria, no event survives. Therefore, the continuum background is neglected.

\section{Branching Fraction Measurement}

\begin{figure}[H]
\centering
\subfigure{\includegraphics[scale=0.40]{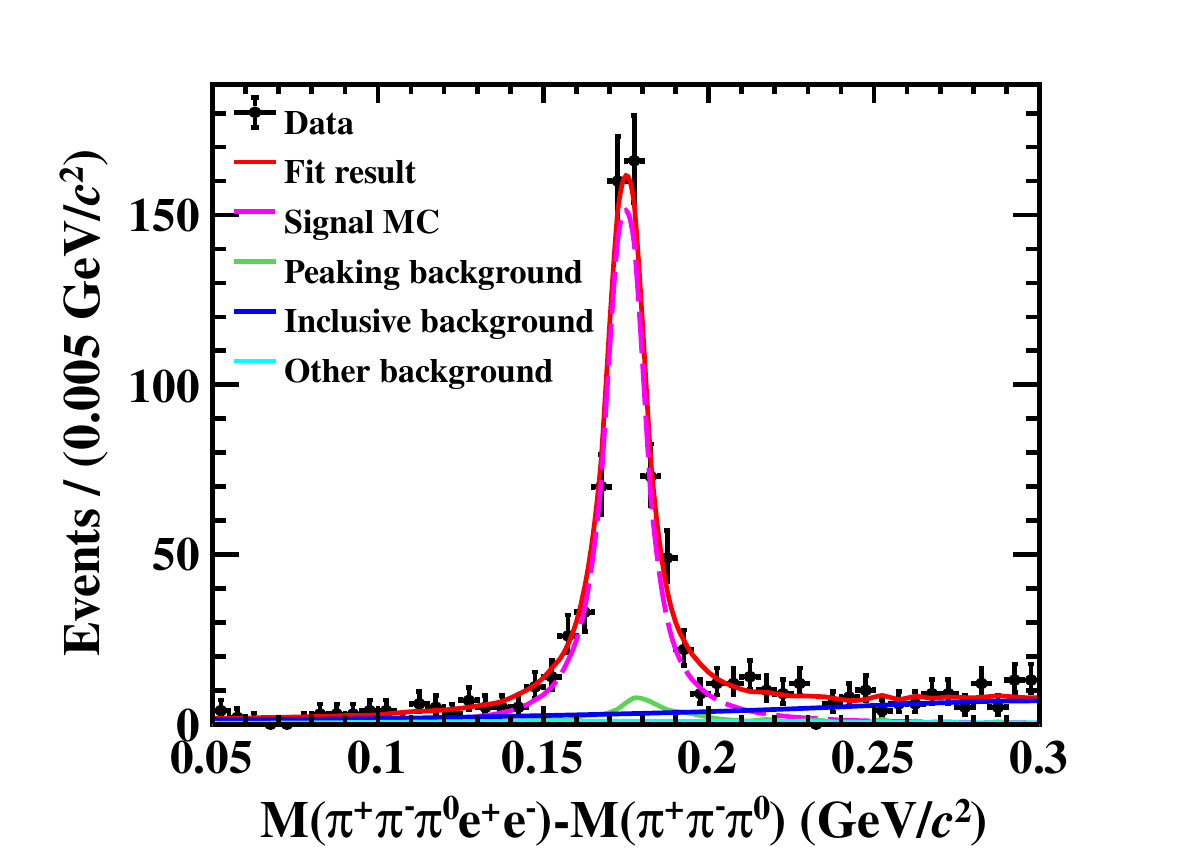}}
\caption{The fit to the distribution of $M(\pi^{+}\pi^{-}\pi^{0}e^{+}e^{-}) - M(\pi^{+}\pi^{-}\pi^{0})$. The dots with error bars represent the data, the red line represents the total fit result and the pink line represents the signal MC sample, the green line represents the peaking background from $J/\psi\to\gamma\eta^{\prime}$, $\eta^{\prime}\to\gamma\omega,\omega\to\pi^+\pi^-\pi^0,\pi^0\to\gamma \gamma$, the blue line represents the background from the inclusive MC sample and the cyan line represents the background from $\jpsi\to\gamma\eta^{\prime}$, $\eta^{\prime}\to\gamma\omega,\omega\to\pi^+\pi^-\pi^0,\pi^0\to\gamma e^+ e^-$.}
\label{fitresult}
\end{figure}

To further exclude the impact of peaking backgrounds on the signal extraction, the number of signal events is determined by fitting the distribution of $M(\pi^+\pi^-\pi^0e^+e^-)-M(\pi^+\pi^-\pi^0)$, which effectively suppress $\eta^{\prime}$ or $\omega$ peaking backgrounds by rendering them smoother.  
To extract the signal yield of $\eta^\prime\rightarrow e^+e^-\omega$, an unbinned maximum-likelihood fit is performed to the distribution of $M(\pi^+\pi^-\pi^0e^+e^-)-M(\pi^+\pi^-\pi^0)$.
The signal probability density function is modeled using the shape obtained from the signal MC simulation.
Since the decay $\eta^{\prime}\to\gamma\omega$ is not included in the inclusive MC sample, exclusive MC samples are generated for $\jpsi\to\gamma\eta^{\prime}$, $\eta^{\prime}\to\gamma\omega$, $\omega\to\pi^+\pi^-\pi^0$, $\pi^0\to\gamma e^+ e^-$.
Accordingly, the non-peaking background is described by the combined shape from the inclusive MC sample and the exclusive MC sample of $J/\psi\to\gamma\eta^{\prime}$, $\eta^{\prime}\to\gamma\omega$, $\omega\to\pi^+\pi^-\pi^0$, $\pi^0\to\gamma e^+ e^-$ .
The peaking background from photon conversion events in $\jpsi\rightarrow\gamma\eta^{\prime}$, $\eta^{\prime}\rightarrow\gamma\omega$, $\omega\to\pi^+\pi^-\pi^0$, $\pi^0\to\gamma \gamma$ process is modeled using the corresponding MC shape.
The fit yields $N_{\mathrm{sig}}= 609.1\pm27.7$ signal events, as shown in Fig.~\ref{fitresult}. The BF of $\eta^\prime \to e^+e^-\omega$ is determined as
\begin{align}
    \mathcal{B}{(\eta^\prime\rightarrow e^{+}e^{-}\omega)}=\frac{N_{\mathrm{sig}}}{N_{J/\psi}\cdot\mathcal{B}_{\rm sub}\cdot\epsilon},
\end{align}
where $N_{J/\psi}$ is the total number of $J/\psi$ events in data, $\epsilon = 7.30\%$ is the detection efficiency, and $\mathcal{B_{\rm sub}}$ denotes the product BFs $\mathcal{B}( \jpsi\rightarrow \gamma\eta^\prime)\cdot\mathcal{B}(\omega\rightarrow \pi^{+}\pi^{-}\pi^{0})\cdot\mathcal{B}(\pi^{0}\rightarrow\gamma\gamma)$, with BFs quoted from the PDG~\cite{ParticleDataGroup:2024cfk}.
The BF of $\eta^\prime\to e^+ e^-\omega$ is calculated to be $(1.79\pm0.09)\times 10^{-4}$, where the uncertainty is statistical only.

\section{Transition Form Factor Measurement}

\begin{figure}[h]
\centering
\subfigure{\includegraphics[scale=0.4]{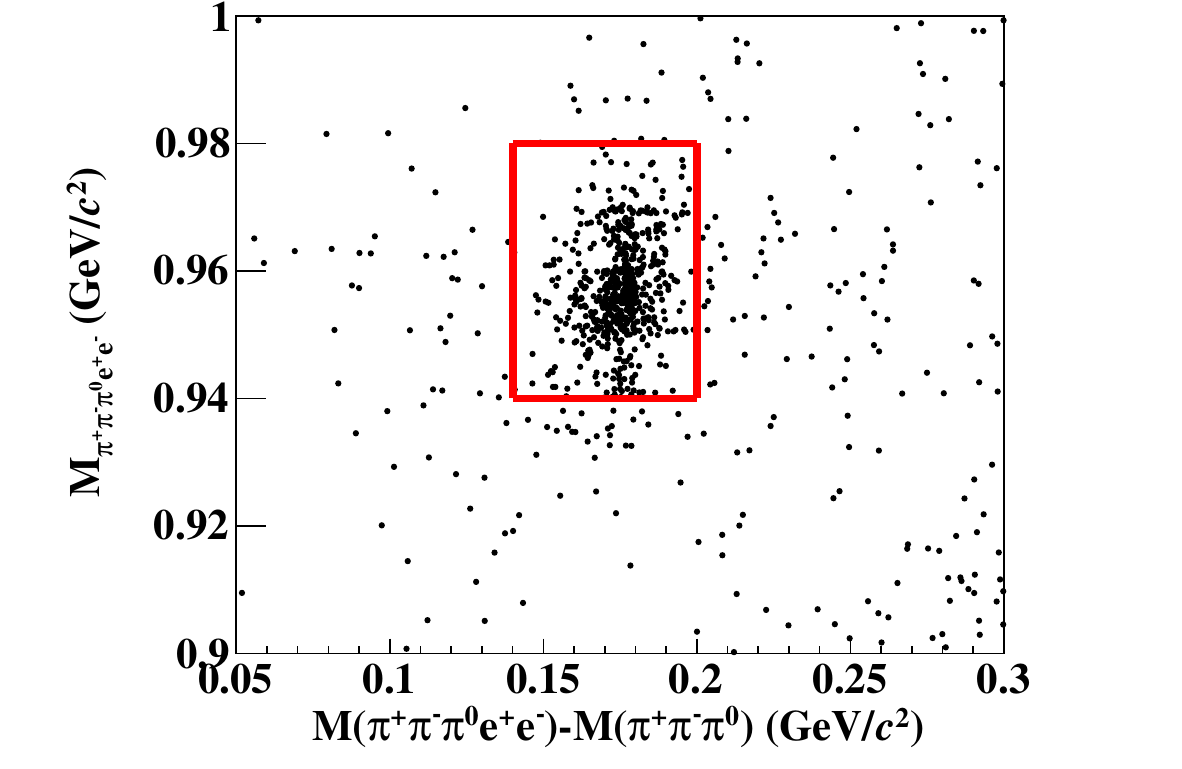}}
\caption{The distribution of $M(\pi^{+}\pi^{-}\pi^{0}e^{+}e^{-})$ versus $M(\pi^{+}\pi^{-}\pi^{0}e^{+}e^{-}) - M(\pi^{+}\pi^{-}\pi^{0})$. The signal region for the TFF measurement is shown in the red square.}
\label{3piee-3pi_3piee}
\end{figure}

To extract the TFF, a cleaner data sample is required. The signal region is further refined as $M(\pi^{+}\pi^{-}\pi^{0}e^{+}e^{-}) \in [0.94, 0.98]~\text{GeV}/c^2$ and $M(\pi^{+}\pi^{-}\pi^{0}e^{+}e^{-}) - M(\pi^{+}\pi^{-}\pi^{0}) \in [0.14, 0.20]~\text{GeV}/c^2$, which enhances the signal purity.
With all the selection criteria applied, 586 events are retained for the TFF analysis. The distribution of $M(\pi^{+}\pi^{-}\pi^{0}e^{+}e^{-})$ versus $M(\pi^{+}\pi^{-}\pi^{0}e^{+}e^{-}) - M(\pi^{+}\pi^{-}\pi^{0})$ is shown in Fig.~\ref{3piee-3pi_3piee}. Background contributions are estimated using the inclusive and exclusive MC samples.

The TFF of the $\eta^\prime$ is extracted by performing an unbinned maximum-likelihood fit on the $e^{+}e^{-}$ invariant mass spectrum of the selected $\eta^\prime\to\omega e^+e^-$ candidates.
The likelihood function is defined as
\begin{align}
\mathcal{L}=\prod_{i=1}^{N}P(\xi_{i}),
\end{align}
where $N$ is the number of observed events, and $\xi_i$ represents the set of four-momenta of the final-state particle for the $i$-th event.
The probability ($P(\xi_{i})$) for observing the $i$-th event is given by
\begin{align}
 P(\xi_{i})=\frac{\omega(\xi_{i})\epsilon(\xi_{i})}{\int d\xi\omega(\xi)\epsilon(\xi)},
 \end{align}
where $\omega(\xi_i) \equiv (d\sigma/d\phi)_i$ represents the differential cross section, $\epsilon(\xi_i)$ the detection efficiency, and $\int d\xi \omega(\xi) \epsilon(\xi)$ represents the measured total cross section.

In this analysis, the squared amplitude for the decay $\eta^\prime\rightarrow e^{+}e^{-}\omega$ is given by Eq.~\ref{eq_a}.

In the log-likelihood calculation, background events are assigned negative weights, i.e.
\begin{align}
-\rm ln\mathcal{L}=\omega^{'}\left[-(\sum_i^{N_{\rm data}}\ln\mathcal{L}_{\rm data}-\sum_j^{N_{\rm bkg}^j}\omega_j\cdot\ln\mathcal{L}_{\rm bkg})\right],
\end{align}
where $i$ runs over all accepted data events, and $j$ runs over the considered background events.
The estimated number of background events, listed in Table~\ref{TFF_BG}, are fixed according to their normalization to the data luminosity.
To obtain an unbiased uncertainty, a normalization factor derived from Ref.~\cite{Langenbruch:2019nwe} is defined as
\begin{align}
\omega^{\prime}=\frac{N_{\rm data}-\sum_j N_{\rm bkg}^j \omega_j}{N_{\rm data}+\sum_j N_{\rm bkg}^j \omega_j^2},
\end{align}
where $N_{\rm data}$ and $N_{\rm bkg}^j$ are the numbers of data events and the $j$-th background component, respectively, and $\omega_j$ is the corresponding background weight.
 \begin{table}[htp]
\begin{center}
\caption{The estimated numbers of background events $N_{\rm bkg}$ for different background components.}
\label{TFF_BG}
\begin{tabular}{c | c }
\hline Component &  $N_{\rm bkg}$    \\
\hline  $\eta^\prime\rightarrow\gamma\omega,\omega\to\pi^+\pi^-\pi^0,\pi^0\to\gamma\gamma$  &$32.8$ \\
\hline  $\eta^\prime\rightarrow\gamma\omega,\omega\to\pi^+\pi^-\pi^0,\pi^0\to\gamma e^+e^-$  &$8.4$ \\
\hline  Inclusive background   &$18.0$     \\
\hline Total  & $59.2$   \\
\hline
\end{tabular}
\end{center}
\end{table}

\begin{figure}[htp]
\begin{center}
\centering
\subfigure{\includegraphics[scale=0.4]{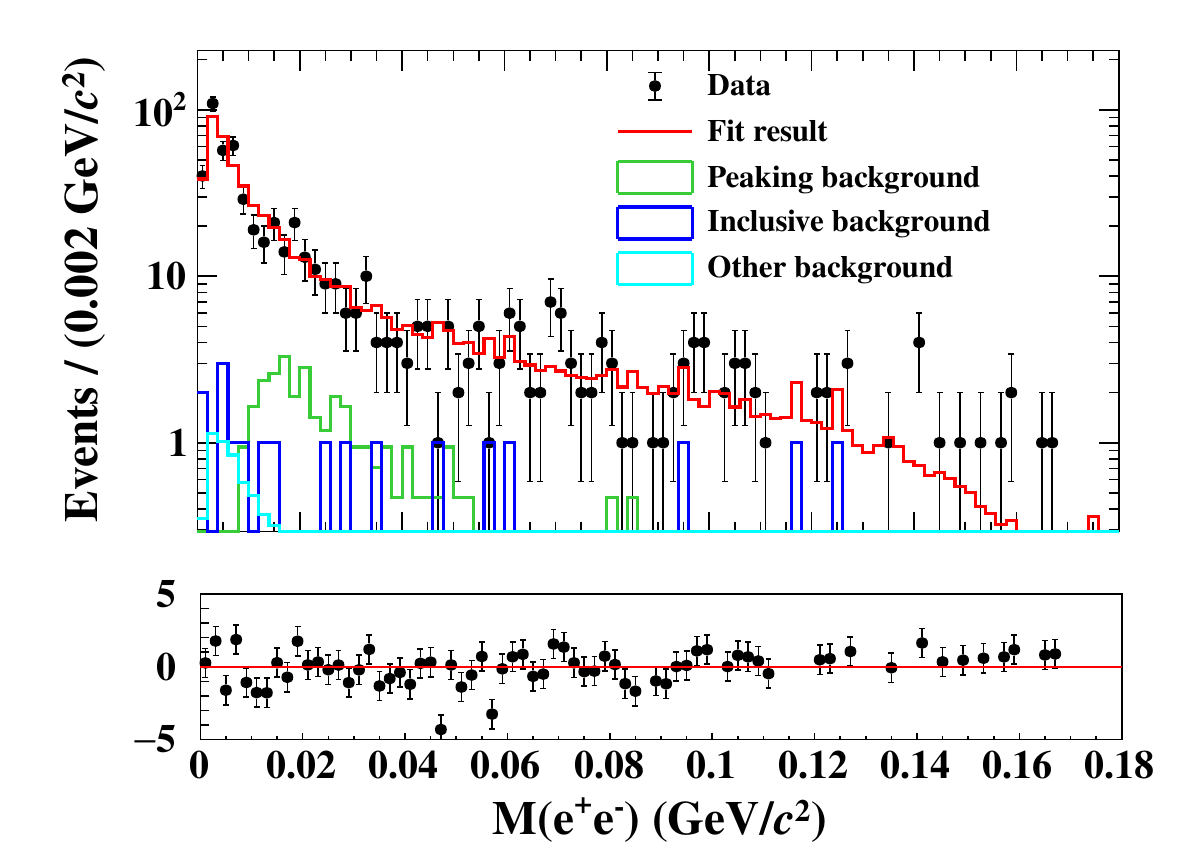}}
\caption{The $e^{+}e^{-}$ invariant mass distribution and the fit result of the TFF. The dots with error bars represent the data, the red line represents the total fit result, the green histogram represents the peaking background from $J/\psi\to\gamma\etap$, $\etap\to\gamma\omega,\omega\to\pi^+\pi^-\pi^0,\pi^0\to\gamma \gamma$, the blue histogram represents the background from the inclusive MC sample and the cyan histogram represents the background from $\jpsi\to\gamma\etap$, $\etap\to\gamma\omega,\omega\to\pi^+\pi^-\pi^0,\pi^0\to\gamma e^+ e^-$. }
\label{FormFactor}
\end{center}
\end{figure}

The fit yields $\Lambda^{-1} = (2.92\pm0.83)\ \text{GeV}^{-1}$, where the uncertainty is statistical only. A comparison between the fit results and the $e^{+}e^{-}$ invariant mass distribution of the data is presented in Fig.~\ref{FormFactor}.

\section{Systematic uncertainties}

The systematic uncertainties of the TFF measurement originate mainly from the MDC tracking and PID of charged particles,  the photon detection, the kinematic fit, the background suppression criteria, and the estimation of background contamination.
The BF measurement shares these uncertainties and suffers from additional effects from the mass fit range, the total number of $J/\psi$ events, and the external BFs from the PDG~\cite{ParticleDataGroup:2024cfk}.
All systematic uncertainties are assumed to be independent and added in quadrature. The individual contributions are summarized in Table~\ref{sys} and are described in detail below.

\begin{itemize}
\item MDC tracking and PID:
The efficiency differences of data and MC simulation for pion MDC tracking and PID are studied using the control sample of $J/\psi\rightarrow\pi^+\pi^-\pi^0$~\cite{Liu:2024uno}.
The tracking efficiencies for electrons are obtained with the control sample of radiative Bhabha scattering $e^+e^-\rightarrow e^+e^-\Gamma_{\rm ISR}$~\cite{Chai:2025xni} (including $J/\psi\rightarrow e^+e^-\gamma_{\rm FSR}$) at $\sqrt{s} = 3.097~\rm{GeV}$.
The differences between data and MC simulation are parametrized as the momentum- and polar-angle-dependent correction factors, which are applied according to the event weights in the signal MC sample.
The weighted uncertainties are taken as the systematic uncertainties.

\item Photon detection: The systematic uncertainty associated with photon reconstruction is obtained with the control sample of $e^+e^-\rightarrow \mu^+\mu^-\gamma_{\rm ISR}$.
The photon detection efficiencies are determined as two-dimensional functions of the photon energy and polar angle.
For the TFF measurement, the efficiency of signal MC simulation is corrected by the ratio $\epsilon_{\rm data}/\epsilon_{\rm MC}$, where $\epsilon_{\rm data}$ and $\epsilon_{\rm MC}$ are the 2D efficiency obtained from the control sample, respectively. The resulting variation in the extracted TFF is assigned as the associated systematic uncertainty.
For the BF measurement, the systematic uncertainties are weighted according to the photon energy and angular distributions in data and MC simulation, and are taken as the corresponding uncertainties.

\item Kinematic fit: To investigate the systematic uncertainty associated with the kinematic fit, the track helix parameter correction method~\cite{BESIII_2013_kmfitfit} is used.
The difference in detection efficiencies between samples with and without the helix parameter correction is taken as the systematic uncertainty.

\item Photon conversion veto: The systematic uncertainty due to the photon conversion veto is estimated by varying the nominal criterion to $R_{xy}>1.9$ cm or $R_{xy}>2.1$ cm.
The resulting change in the final results is assigned as the associated uncertainty.

\item $\pi^0$ or $\eta^\prime$ mass window: The systematic uncertainty arising from the $\pi^0$ or $\eta^\prime$ mass window is determined by varying the corresponding mass window by one standard deviation. The resulting differences of the BF and the TFF are taken as the uncertainty.

\item $M(\gamma_{1/3} e^+e^-)$ range: The uncertainty arising from the $M(\gamma_{1/3} e^+e^-)$ range is determined by changing the $M(\gamma_{1/3} e^+e^-)$ range by $0.02~\mathrm{GeV}/c^2$ and $0.008~\mathrm{GeV}/c^2$, respectively. The corresponding changes in the BF and the TFF are taken as the systematic uncertainties.

\item Peaking background: The peaking background contribution is estimated using the BFs of $ \jpsi\rightarrow \gamma\eta^\prime$, $\eta^\prime\rightarrow \gamma\omega$, $\omega\rightarrow \pi^{+}\pi^{-}\pi^{0}$, $\pi^{0}\rightarrow\gamma\gamma$ quoted from the PDG~\cite{ParticleDataGroup:2024cfk}. By varying the BFs within their $\pm 1\sigma$ of uncertainties, the change in the final result is taken as the corresponding uncertainty.

\item Non-peaking background: The non-peaking background contribution is estimated using the BFs of $ \jpsi\rightarrow \gamma\eta^\prime$, $\eta^\prime\rightarrow \gamma\omega$, $\omega\rightarrow \pi^{+}\pi^{-}\pi^{0}$, $\pi^{0}\rightarrow\gamma e^+e^-$ quoted from the PDG~\cite{ParticleDataGroup:2024cfk}. The number of background events is varied by $\pm1\sigma$, and the resulting change in the final results is taken as the systematic uncertainty.

\item Fit range: The uncertainty arising from the fit range in the BF is determined to be 2.2\% by changing the fit range by $\pm~0.02 ~\mathrm{GeV}/c^2$.

\item Number of $J/\psi$ events: The number of $J/\psi$ events is determined to be $(10087\pm44)\times10^6$ from the inclusive hadron events, corresponding to a relative uncertainty of 0.4\%~\cite{BESIII_2021_njpsi}.

\item  Intermediate decay: The uncertainties of the quoted BFs for the intermediate decays $J/\psi \to \gamma\eta^\prime$, $\omega\pi^+\pi^-\pi^0$, and $\pi^0\to\gamma\gamma$ are taken from the PDG~\cite{ParticleDataGroup:2024cfk}.

\end{itemize}

\begin{table}[ht]
\begin{center}
\caption{Summary of the systematic uncertainties for BF and TFF measurements (in percent). }
\label{sys}
\begin{tabular}{l | c | c}
\hline  Source & BF  & ~TFF~ \\
\hline  MDC tracking         & 5.7    &0.7\\
          PID                & 1.1    &0.3\\
 Photon detection            & 1.2    &0.3\\
    Kinematic fit            & 0.6    &1.0\\
   Photon conversion veto    & 0.6    &1.4\\
   $\pi^{0}$ mass window     & 0.6    &3.1 \\
  $\eta^\prime$ mass window  & 1.1    &3.1\\
$M(\gamma_{1}e^+e^-)$ range     & 0.1    &0.3\\
$M(\gamma_{3}e^+e^-)$ range     & 0.6    &0.7\\
 Peaking background          & 1.1    &1.0\\
Non-peaking background       & 0.6    &0.7 \\
    Fit range            & 2.2    &-\\
Number of $J/\psi$ events     & 0.4    &-\\
  Intermediate decays         & 1.4    &-\\
\hline Total                  & 6.8  &5.0 \\
\hline
\end{tabular}
\end{center}
\end{table}

\section{SUMMARY}
\begin{figure}[H]
\centering
\subfigure{\includegraphics[scale=0.40]{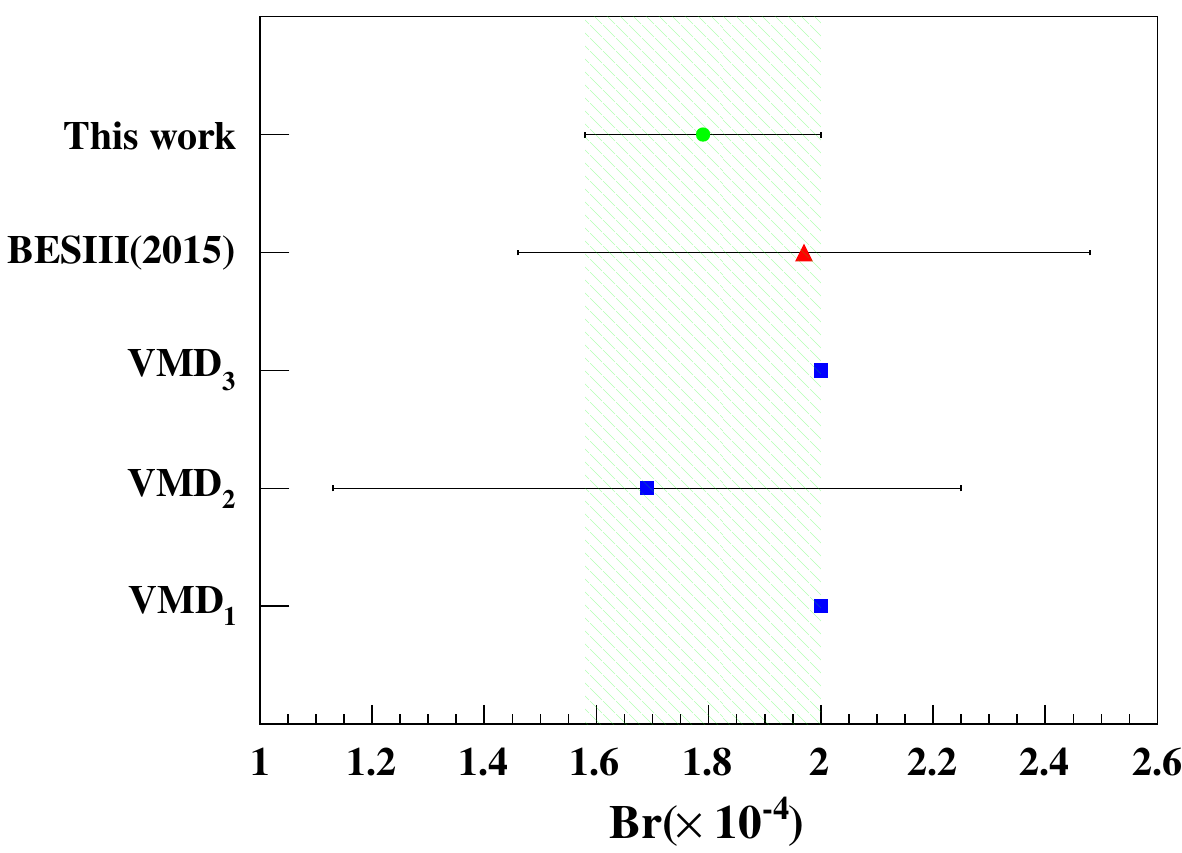}}
\caption{The BFs of the $\eta^{\prime}\to e^+e^-\omega$ from different theoretical predictions~\cite{Faessler:1999de,Terschlusen:2012xw,Yang:2014hwa} (blue squares), the previous experiments~\cite{BESIII:2015jiz} (red triangle), and the results of this work (green squares and green dot).}
\label{br}
\end{figure}

Based on a sample of $(10087\pm44)\times 10^6$ $J/\psi$ events collected with the BESIII detector, the decay $\eta^\prime\to e^+e^-\omega$ is studied.
The BF is determined to be $\mathcal{B}(\eta^\prime\rightarrow e^{+}e^{-}\omega ) = (1.79\pm0.09~(\rm stat.)\pm0.12~(\rm syst.))\times10^{-4}$. This result is consistent with the previous measurement and theoretical predictions, and improves the experimental precision by a factor of $2.7$, as shown in Fig.~\ref{br}.

The transition form factor parameter $\Lambda^{-1}$ of $\eta^\prime\rightarrow e^{+}e^{-}\omega$  is measured for the first time to be $(2.92\pm0.83~(\rm stat.)\pm0.15~(\rm syst.))~ \mathrm{GeV^{-1}}$. This parameter provides important input for phenomenological studies of $\eta^{\prime}$ electromagnetic transitions. Since the present measurement is limited by statistics, a larger $\eta^{\prime}$ data sample is needed to further improve the precision in the future.

\begin{acknowledgments}
The BESIII Collaboration thanks the staff of BEPCII~\cite{BEPCII} and the IHEP computing center for their strong support. This work is supported in part by National Key R\&D Program of China under Contracts No. 2025YFA1613900, No. 2023YFA1606000, No. 2023YFA1606704; National Natural Science Foundation of China (NSFC) under Contracts No. 12225509, No. 12465015, No. 12575087, No. 11635010, No. 11935015, No. 11935016, No. 11935018, No. 12025502, No. 12035009, No. 12035013, No. 12061131003, No. 12192260, No. 12192261, No. 12192262, No. 12192263, No. 12192264, No. 12192265, No. 12221005, No. 12235017, No. 12342502, No. 12361141819; the Chinese Academy of Sciences (CAS) Large-Scale Scientific Facility Program; the Strategic Priority Research Program of Chinese Academy of Sciences under Contract No. XDA0480600; CAS under Contract No. YSBR-101; 100 Talents Program of CAS; The Institute of Nuclear and Particle Physics (INPAC) and Shanghai Key Laboratory for Particle Physics and Cosmology; ERC under Contract No. 758462; German Research Foundation DFG under Contract No. FOR5327; Istituto Nazionale di Fisica Nucleare, Italy; Knut and Alice Wallenberg Foundation under Contracts No. 2021.0174, No. 2021.0299, No. 2023.0315; Ministry of Development of Turkey under Contract No. DPT2006K-120470; National Research Foundation of Korea under Contract No. NRF-2022R1A2C1092335; National Science and Technology fund of Mongolia; Polish National Science Centre under Contract No. 2024/53/B/ST2/00975; STFC (United Kingdom); Swedish Research Council under Contract No. 2019.04595; U. S. Department of Energy under Contract No. DE-FG02-05ER41374

\end{acknowledgments}

\bibliography{apssamp}

\end{document}

%% file: authorlist_2025-12-30.tex
M.~Ablikim$^{1}$\BESIIIorcid{0000-0002-3935-619X},
M.~N.~Achasov$^{4,c}$\BESIIIorcid{0000-0002-9400-8622},
P.~Adlarson$^{83}$\BESIIIorcid{0000-0001-6280-3851},
X.~C.~Ai$^{89}$\BESIIIorcid{0000-0003-3856-2415},
C.~S.~Akondi$^{31A,31B}$\BESIIIorcid{0000-0001-6303-5217},
R.~Aliberti$^{39}$\BESIIIorcid{0000-0003-3500-4012},
A.~Amoroso$^{82A,82C}$\BESIIIorcid{0000-0002-3095-8610},
Q.~An$^{79,65,\dagger}$,
Y.~H.~An$^{89}$\BESIIIorcid{0009-0008-3419-0849},
Y.~Bai$^{63}$\BESIIIorcid{0000-0001-6593-5665}
% ...
\\
\vspace{0.2cm}
(BESIII Collaboration)\\
\vspace{0.2cm} {\it
$^{1}$ Institute of High Energy Physics, Beijing 100049, People's Republic of China\\
% ...
$^{4}$ Budker Institute of Nuclear Physics SB RAS (BINP), Novosibirsk 630090, Russia\\
% ...
$^{31}$ INFN Sezione di Ferrara, (A)INFN Sezione di Ferrara, I-44122, Ferrara, Italy; (B)University of Ferrara, I-44122, Ferrara, Italy\\
% ...
$^{39}$ Johannes Gutenberg University of Mainz, Johann-Joachim-Becher-Weg 45, D-55099 Mainz, Germany\\
% ...
$^{63}$ Southeast University, Nanjing 211100, People's Republic of China\\
% ...
$^{65}$ State Key Laboratory of Particle Detection and Electronics, Beijing 100049, Hefei 230026, People's Republic of China\\
% ...
$^{79}$ University of Science and Technology of China, Hefei 230026, People's Republic of China\\
% ...
$^{82}$ University of Turin and INFN, (A)University of Turin, I-10125, Turin, Italy; (B)University of Eastern Piedmont, I-15121, Alessandria, Italy; (C)INFN, I-10125, Turin, Italy\\
$^{83}$ Uppsala University, Box 516, SE-75120 Uppsala, Sweden\\
% ...
$^{89}$ Zhengzhou University, Zhengzhou 450001, People's Republic of China\\

\vspace{0.2cm}
$^{\dagger}$ Deceased\\
% ...
$^{c}$ Also at the Novosibirsk State University, Novosibirsk, 630090, Russia\\
% ...

}